# Recycling of SmCo$_5$ magnets by HD process

**Anas Eldosouky**[1,2], **Irena Škulj**[1]

[1]*Magneti Ljubljana, d.d., Stegne 37, 1000 Ljubljana, Slovenia*

[2]*Jožef Stefan International Postgraduate School, Jamova cesta 39, 1000 Ljubljana, Slovenia*

**Abstract**

Hydrogen decrepitation process has been applied for the first time for the direct recycling of SmCo$_5$ magnets. Industrially produced sintered SmCo$_5$ magnets were decrepitated by hydrogen gas at a pressure of 1 bar to 9.5 bar at room temperature in a planetary rotating jar. After decrepitation, the starting sintered magnets were reduced to a powder with a particle size of less than 200 μm. The produced powder was used for the preparation of recycled SmCo$_5$ magnets. Scanning electron microscopy, energy-dispersive X-ray spectroscopy, X-ray diffraction studies and magnetic measurements were used to follow the decrepitation and the sintering processes. The measured remanence and maximum energy product of the recycled magnet are 0.945 T and 171.13 kJ/m$^3$, respectively, in comparison with 0.915 T and 156.92 kJ/m$^3$, respectively for the original magnet before recycling. It was also observed that, there is refinement in the microstructure after recycling in comparison to the original magnet.

1. Introduction

*Corresponding author. Address: Stegne 37, 1000 Ljubljana, Slovenia. Tel.: +386 1 511 13 04. E-mail address: Anas.Eldosouky@magneti.si (A. Eldosouky).

China has the largest reserves of rare earth elements (REEs) in the world and it dominates the market of REEs production. Around 85% of REEs are produced in China. In 2010 and 2014, the European Commission published two reports identifying the critical raw materials for the European Union, in which REEs were the most problematic in terms of supply risk and very valuable in terms of economic importance.[1,2] Three solutions were discussed for the REEs supply risk issue; recycling of REEs containing materials, REEs substitution or to open new REEs mines outside China. Among those options, recycling is the most appropriate in terms of time, environmental aspects and cost.

REE magnet industry consumes about 20% of the worldwide produced REEs.[3,4] This makes REE magnets as primary candidates for the REEs recycling. Nowadays, NdFeB magnets represent more than 90% of the REE permanent magnets production. But, SmCo magnets have much higher coercivity and better temperature resistance, and many experts expect that, the use of SmCo will grow in the next years, especially because the presence of very expensive Dy content with the highest supply risk in the NdFeB magnets; in order to enhance the magnetic properties.[5] Therefore, innovative direct and indirect recycling methods should be established for the $SmCo_5$ magnets.

Hydrogen decrepitation (HD) can be described as a size reduction technique, where the material under investigation reacts with hydrogen gas under pressure and temperature. The hydride form of the material or one of its phases is usually formed due to the HD process. Moreover, the process is usually accompanied by volume expansion due to the hydrogen absorption in the interstitial sites of the magnet's phases(s) and the disintegration of the material to lower particle sizes. HD is a well-established technique for the production and recycling of NdFeB magnets at relatively low hydrogen pressures of less than 10 bar and room temperature.[6-8]

Previous studies have shown that, in contrast to NdFeB, high conditions of pressure and/or temperature are usually required to induce hydrogen absorption of $SmCo_5$. Zijlstra et al.[9] showed that, under a hydrogen pressure of 20 bar; the hexagonal $SmCo_5$ absorbs 2.5 moles of hydrogen gas and converts to orthorhombic structure. The starting hexagonal structure is recovered when the hydrogen gas pressure is relieved. Similarly, van Vucht et al.[10] showed that hydrogenation under 100 bar for 2 days results in the formation of orthorhombic $SmCo_5H_3$ with 10 vol.% unit cell expansion of the hexagonal $SmCo_5$.

The conditions required for the decrepitation depend on many variables as the chemical composition, the microstructure, the particle size, the pressure cycling.[6,11-13] Harris[6] has mentioned that, the decrepitation of $SmCo_5$ compounds starts with intergranular absorption of hydrogen by the Sm-rich phases which is then followed by transgranular fracture of the matrix $SmCo_5$ phase. He showed that, the alloy with a composition of $SmCo_{4.8}$ reacts at lower pressure than the one required for the alloy with a composition of $SmCo_5$; due to the higher content of the Sm-rich phases in the previous alloy which have higher hydrogen absorption affinity than the $SmCo_5$ phase and therefore, enhance the hydrogen decrepitation. Similar behavior was observed by Raichlen et al.[12], they showed that, alloys of composition of 31.3 wt% Sm and 68.7 wt% Co do not decrepitate even under 100 bar for 2 days and the Sm content in the alloy has to be higher than 31.1 wt% for the material to absorb hydrogen under the latter conditions. Handstein et al.[13] showed that, the pressure required for the interstitial absorption can be decreased with increasing the applied temperature. The hydrogen absorption peak for the $SmCo_5$ alloy was at 10 bar, 120 °C or 40 bar, 103 °C.

Szmaja et al.[14] and Riley[15] showed that the microstructure of the commercially produced $SmCo_5$ magnets usually consists of the matrix $SmCo_5$ phase of a grain size up to 30 µm, randomly distributed mixture of $Sm_2Co_7$ and $Sm_5Co_{19}$ in the size range of 5 μm to 50 μm, other Sm-rich phases and Sm-oxides of smaller sizes and pores.

In this paper, sintered $SmCo_5$ magnets were hydrogen decrepitated under 1-9.5 bar, room temperature in a rotating equipment. The rotation force was used to decrease the conditions required for the HD process. The produced powder was re-sintered again to produce recycled magnets having higher magnetic properties in comparison to the original magnet.

## 2. Material and methods

Sintered $SmCo_5$ magnets of cylindrical shape with a chemical composition of 36.30 wt% Sm, 63.22 wt% Co, 0.37 wt% O, were used for the decrepitation experiments. The volume of the magnet was around 0.4 cm³. 200 g of the magnet was put in a 100 mL stainless steel jar with a lid, which enables evacuation and gas filling up to a pressure of 15 bar (Fig. 1). After evacuation, the jar was filled with hydrogen gas of purity >99.999% to different pressures of 1 bar, 2.5 bar, 4 bar and 9.5 bar. The jar was rotated in a planetary ball mill for 3h at 250 rpm to perform the decrepitation reaction. Every 15 minutes, the rotation was stopped and the jar was filled with hydrogen gas to the starting pressure; to compensate for the hydrogen absorbed by the material. The produced powder after decrepitation was taken inside a glove box with oxygen concentration less than 10 ppm. The decrepitated powder was then jet milled in nitrogen atmosphere to a particle size of less than 20 μm. The milled powder was used to produce anisotropic magnets by means of conventional sintering after isostatic pressing. The sintering profile for the milled decrepitated powder is shown in Fig. 2.

The demagnetization curves for the recycled and original magnets were measured by Magnet-Physik's permagraph of applied field up to 1800 kA/m. PANalytical Empyrean multi-purpose diffractometer was used to analyse the X-ray diffraction (XRD) patterns of the original magnet and the decrepitated powders produced by different pressures. Helios NanoLab™ 650 scanning electron microscope was used to study the microstructure of the decrepitated powders and the original and the recycled magnets. Oxygen content in the original and recycled magnets was measured by Eltra ON 900 oxygen/nitrogen determinator. The density of the sintered sample were measured by an Archimedes' principle based equipment.

A mixture of chromium oxide (4g) and sodium sulfate (1g) in 50 mL deionized water was used to etch the sintered magnets for the microstructure studies.

**Figure 1.**

**Figure 2.**

3. **Results and discussion**

The secondary electron (SE) images of the powders after decrepitation produced by different pressures at 1 bar, 2.5 bar, 4 bar and 9.5 bar are shown in Fig. 3. For all pressures, there is a large distribution of particle sizes, with the presence of very fine particles, lower than 300 nm. The upper particle size of the decrepitated powder throughout the samples was less than 200 μm. By definition, HD is always accompanied by the presence of cracks in the decrepitated powders due to hydrogen absorption in the microstructure of the material. The presence of the rotation force had the effect of decreasing these cracks and the production of smaller particle sizes, however, some cracks were still visible in the powders.

**Figure 3.**

Fig. 4 shows the XRD patterns for the starting magnet and the decrepitated powders. It is shown that the patterns for the magnet and the powders after decrepitation by different pressures are very similar. The HD reaction starts by the hydrogen reaction with the $Sm_2Co_7$ and other Sm-rich phases of low concentration, with the formation of Sm-hydrides and Co-rich phases. This is confirmed by the disappearance of the $Sm_2Co_7$ peak in the decrepitated powders. Sm-hydrides could not be identified because of the low concentration and the oxidation to form Sm-oxides during the preparation and the measurement steps. Hydrogen is also absorbed in the interstitial sites of the $SmCo_5$ matrix phase. As described earlier, when the hydrogen pressure above the magnet is relived, the interstitially absorbed hydrogen is relived from the $SmCo_5$ structure. The decrease in the intensity of some peaks of the $SmCo_5$ phase after decrepitation indicates that in addition to the interstitial hydrogen absorption, a small portion of $SmCo_5$ is dissociated by hydrogen gas to form Sm-hydrides and Co-rich phases.

**Figure 4.**

As the XRD patterns and the microstructures were similar for all the decrepitated powders from different pressures, it is expected that the magnets produced using any powder from any pressure will have the same properties. The powder produced from decrepitation by 4 bars was selected for further sintering experiments.

After conventional sintering, the measured density for the recycled magnet made from the milled decrepitated powder was 8.33 g/cm³, which is slightly higher than the density of the original magnet of 8.28 g/cm³. For comparison, Fig. 5 shows the back scattered electron (BSE) images of the original magnet before decrepitation and the recycled magnet. The microstructure of the original magnet has been recovered after recycling. By using energy-dispersive X-ray spectroscopy (EDX) analysis, it was measured that, both the original and the recycled magnet consist of the $SmCo_5$ matrix phase, the $Sm_2Co_7$ phase shown as light gray regions distributed randomly along the matrix phase, small brightest regions of Sm-rich phases and Sm-oxides. During re-sintering process, the oxygen content has been increased from 0.37 wt% in the original magnet to 0.48 wt% in the recycled magnet. Fig. 6 shows the SE images of the original and recycled magnets after etching. After recycling, there is a refinement in the microstructure. This can be explained due to the presence of nano-size particles in the recycled powders after decrepitation which are not present in the powder used to prepare the original magnet. The original magnet was produced by a non-decrepitated powder of a particle size of 4-20 µm.

**Figure 5.**

**Figure 6.**

Fig. 7 shows the demagnetization curves for the original and the recycled magnets. The remanence ($B_r$) for the recycled magnets was 0.945 T in comparison to 0.915 T for the original magnet. The increased in $B_r$ may be explained as the presence of nano-particle size in the recycled powders increases the alignment and compaction of the powder which results in an increase in $B_r$, this is also confirmed by the microstructure refinement and the increase in density of the recycled magnet. Moreover, the recycled magnet was compacted by isostatic pressing, where the original magnet was prepared by axial pressing. The use of isostatic pressing is known to be a reason for the $B_r$ enhancement when compared to axial pressing. As a result of $B_r$ enhancement, the maximum energy product of the recycled magnet was 171.13 kJ/m$^3$ in comparison to 156.92 kJ/m$^3$ for the original magnet.

**Figure 7.**

4. **Conclusion**

Hydrogen decrepitation was successfully used for the recycling of SmCo$_5$ sintered magnets. Hydrogen pressure as low as atmospheric pressure at room temperature can be used for the production of decrepitated powders, which are further milled to a particle size of less than 20 μm. The milled powders are used for the production of new sintered magnets using the conventional sintering route with an improvement in magnetic properties after recycling.

**Acknowledgement**

The research leading to these results has received funding from the European Community's Horizon 2020 Programme ([H2020/2014-2019]) under Grant Agreement no. 674973 (MSCA-ETN DEMETER). This publication reflects only the authors' view, exempting the Community from any liability. Project website: http://etn-demeter.eu/.

**Figure captions**

**Figure 1.** Stainless steel jar with a lid, used for the decrepitation process

**Figure 2.** The sintering profile for the milled decrepitated $SmCo_5$ powder. Sintering temperature; 1170 °C; heat treatment temperature, 880 °C.

**Figure 3.** SE images of the decrepitated powder of sintered $SmCo_5$ magnet, the applied pressures were 1 bar, 2.5 bar, 4 bar and 9.5 bar for (a), (b), (c) and (d), respectively.

**Figure 4.** XRD patterns for the original magnet and the decrepitated powders by 1-9.5 bar. All the unmarked peaks related to the $SmCo_5$ phase. The $Sm_2Co_7$ phase has been disappeared after decrepitation with the appearances of new phases of Sm-hydrides, Sm-oxides and Co-rich phases.

**Figure 5.** BSE images for the original magnet (left) and the recycled magnet (right). Both magnets consists of $SmCo_5$ matrix phase, $Sm_2Co_7$ light gray clusters with bright regions of Sm-rich phases and Sm-oxides, and pores.

**Figure 6.** SE images for the etched original magnet (left) and the recycled magnet (right). In comparison to the original magnet, there is microstructure refinement for the magnet after recycling.

**Figure 7.** The demagnetization curves for the original and the recycled magnets by applying a reverse magnetic field H.

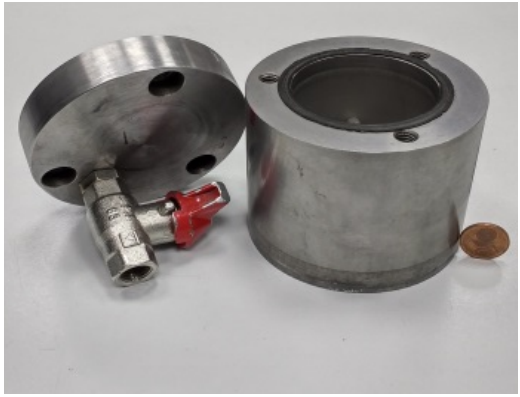

Figure 1. (coloured)

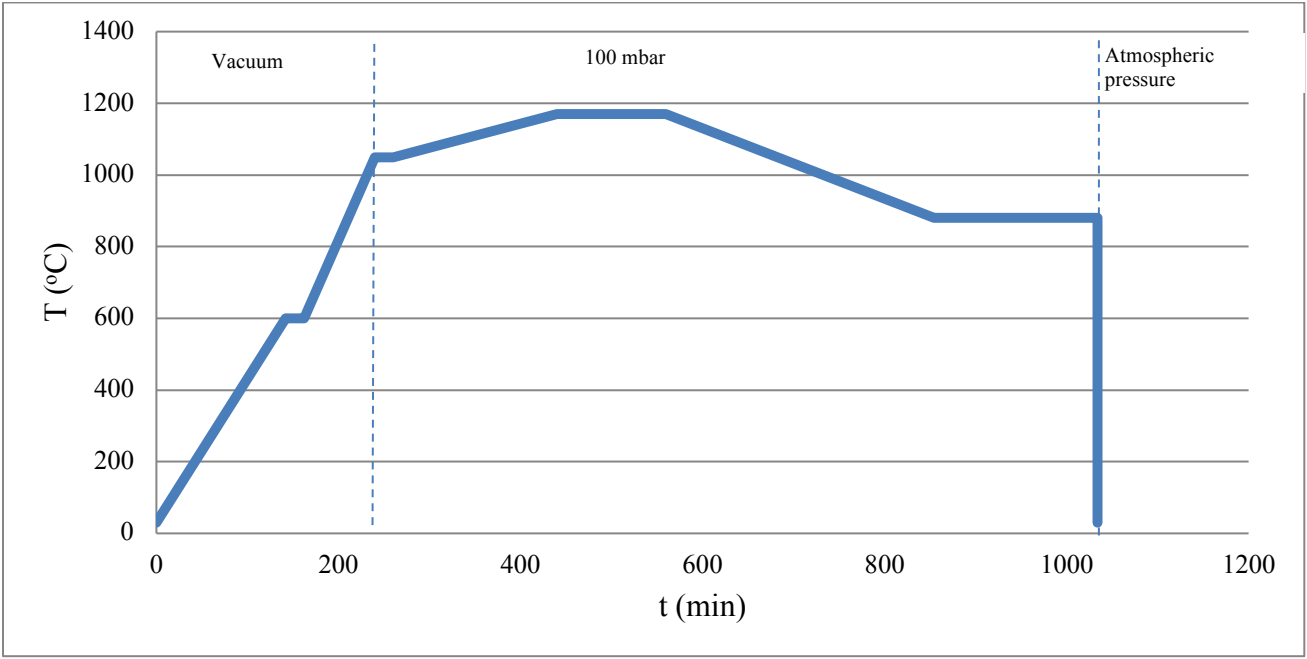

Figure 2.

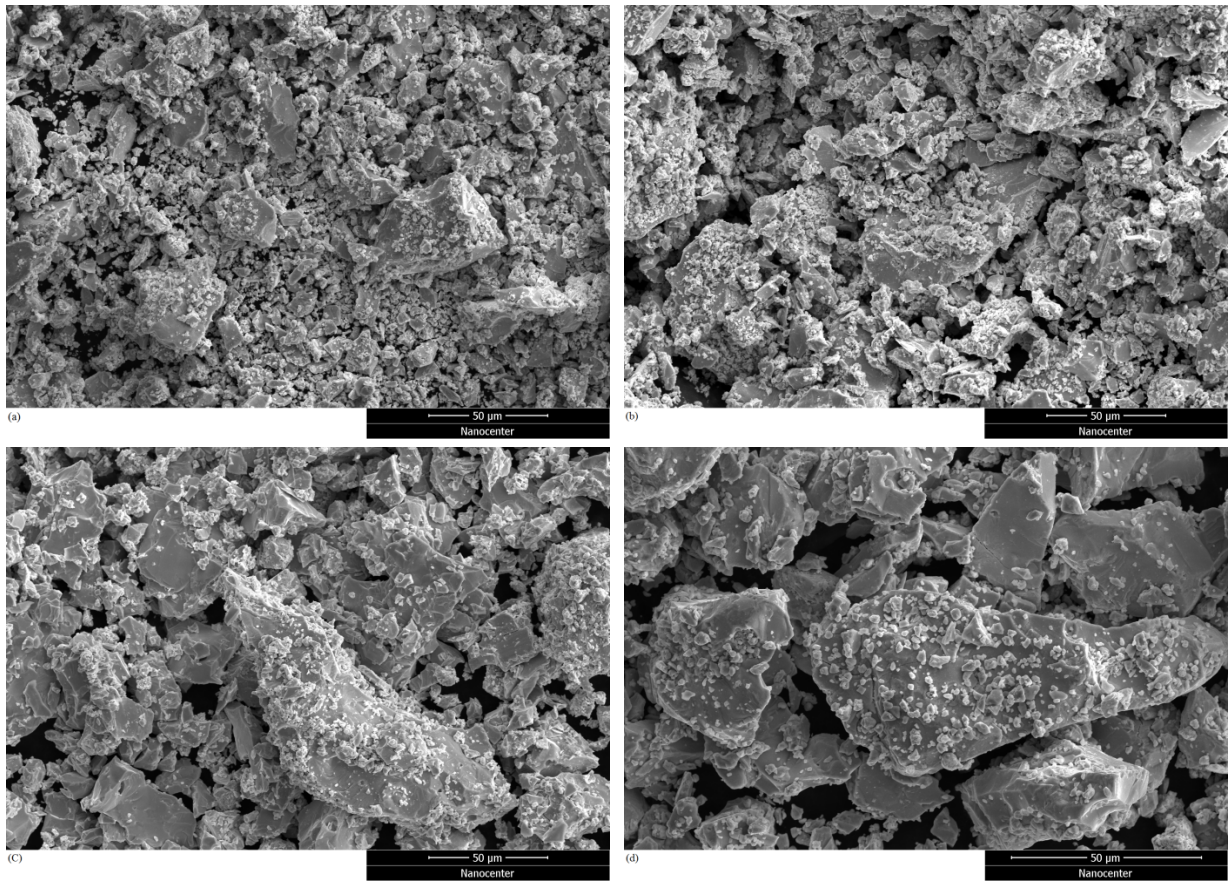

Figure 3.

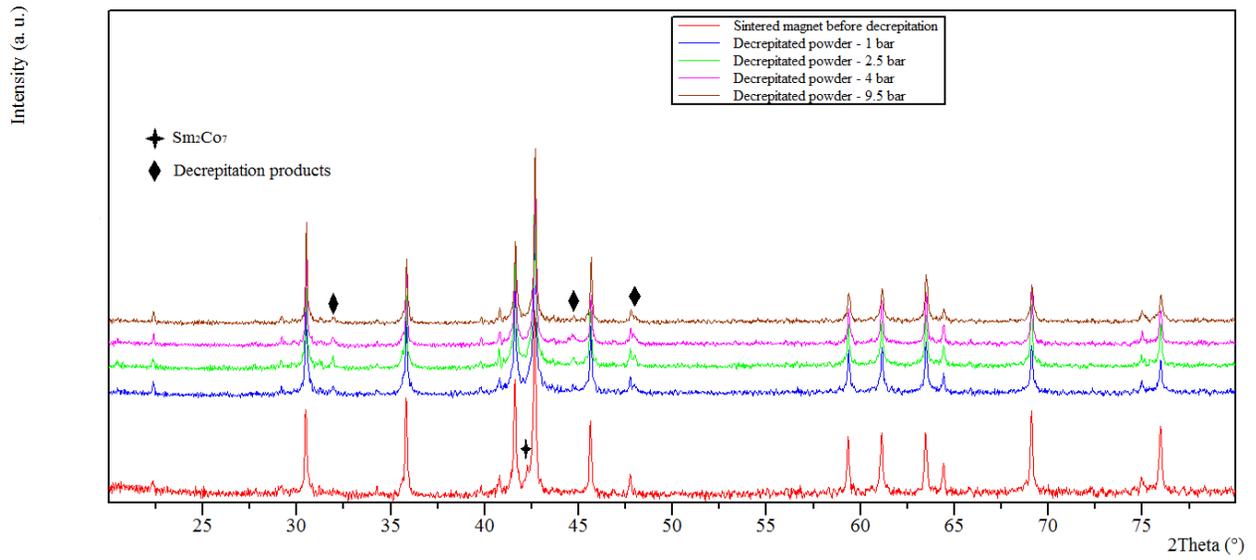

Figure 4. (coloured)

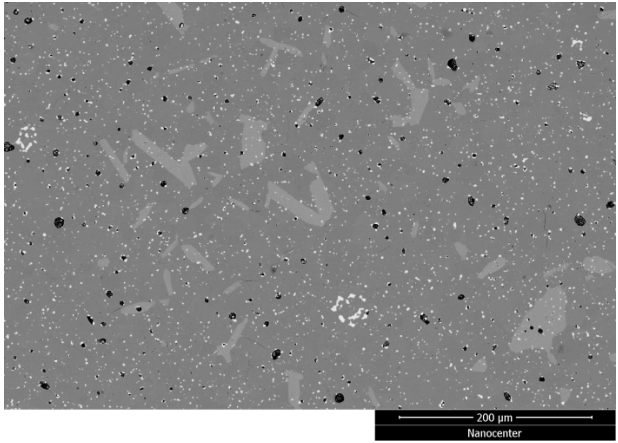 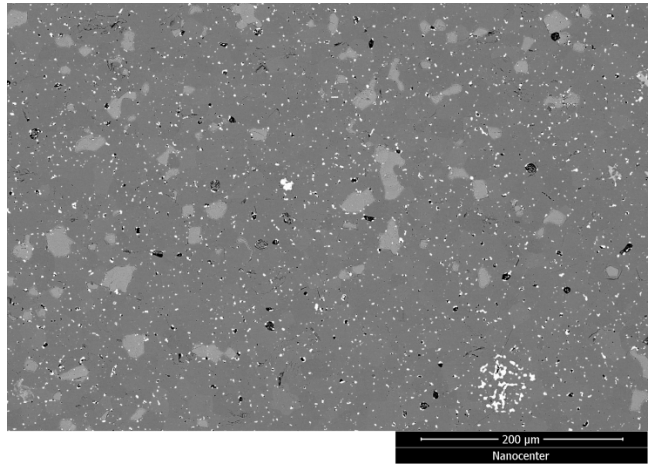

Figure 5.

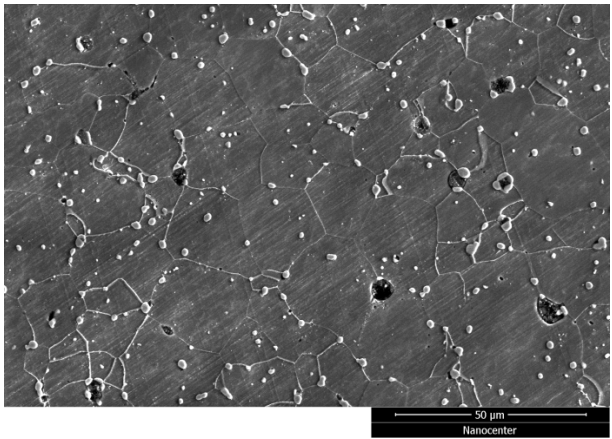 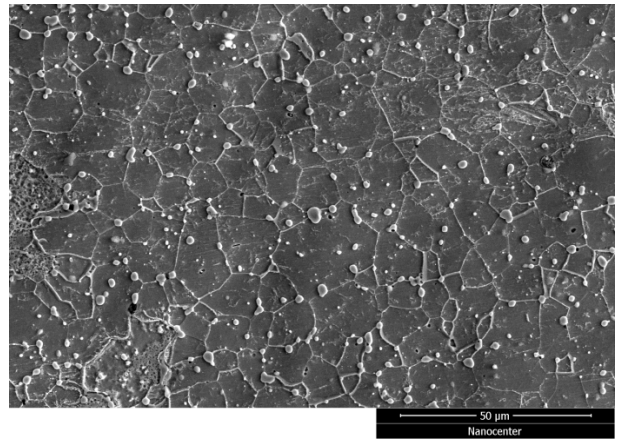

Figure 6.

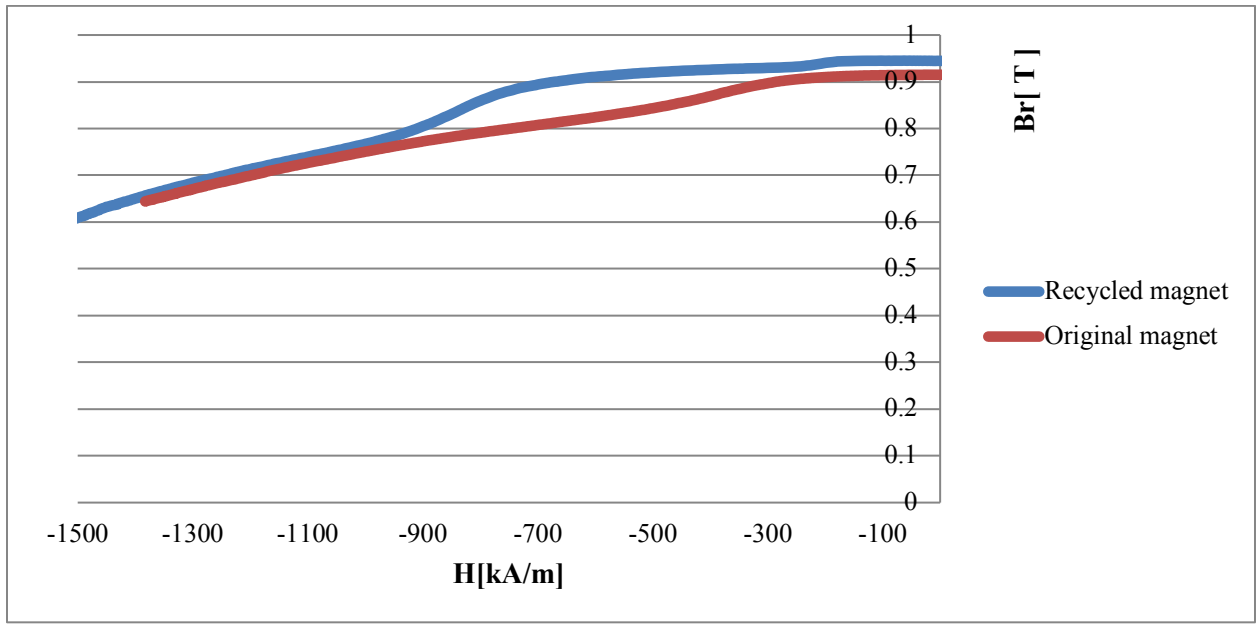

Figure 7. (coloured)